# Low-Symmetry Monoclinic Phases and Polarization Rotation Path Mediated By Epitaxial Strain in Multiferroic $BiFeO_3$ Thin Films

*By Zuhuang Chen, Zhenlin Luo, Chuanwei Huang, Yajun Qi, Ping Yang, Lu You, Chuansheng Hu, Tom Wu, Junling Wang, Chen Gao, Thirumany Sritharan, and Lang Chen\**

[*]   Prof. L. Chen, Z. H. Chen, C. W. Huang, Y. J. Qi, L. You, Prof. J. L. Wang, Prof. T. Sritharan

School of Materials Science and Engineering, Nanyang Technological University

Singapore, 639798 (Singapore)

E-mail: langchen@ntu.edu.sg

Dr. Z. L. Luo, C. S. Hu, Prof. C. Gao

National Synchrotron Radiation Laboratory & Department of Materials Science and Engineering, University of Science and Technology of China

Hefei, 230029 (China)

Dr. P. Yang

Singapore Synchrotron Light Source (SSLS), National University of Singapore

5 Research Link, 117603 (Singapore)

Prof. T. Wu

Division of Physics and Applied Physics, School of Physical and Mathematical Sciences

Nanyang Technological University

Singapore, 637371 (Singapore)

Keywords: strain, multiferroic, $BiFeO_3$, monoclinic phase, morphotropic phase boundary

**Abstract**

A morphotropic phase boundary driven by epitaxial strain has been observed in a lead-free multiferroic BiFeO$_3$ thin films and the strain-driven phase transitions were widely reported to be iso-symmetric *Cc-Cc* ones by recent works. In this paper, we suggest that the tetragonal-like BiFeO$_3$ phase identified in epitaxial films on (001) LaAlO$_3$ single crystal substrates is monoclinic $M_C$. This $M_C$ phase is different from $M_A$ type monoclinic phase reported in BiFeO$_3$ films grown on low mismatch substrates, such as SrTiO$_3$. This is confirmed not only by synchrotron x-ray studies but also by piezoresponse force microscopy measurements. The polarization vectors of the tetragonal-like phase lie in the (100) plane, not the (1$\bar{1}$0) plane as previously reported. A phenomenological analysis was proposed to explain the formation of $M_C$ Phase. Such a low symmetry $M_C$ phase, with its linkage to $M_A$ phase and the multiphase coexistence open an avenue for large piezoelectric response in BiFeO$_3$ films and shed light on a complete understanding towards possible polarization rotation paths and enhanced multiferroicity in BiFeO$_3$ films mediated by epitaxial strain. This work may also aid the understanding of developing new lead-free strain-driven morphotropic phase boundary in other ferroic systems.

## 1. Introduction

Monoclinic phases have been reported in lead-based ferroelectrics, such as $Pb(Zr_{1-x}Ti_x)O_3$,[1] $(1-x)\ Pb(Zn_{1/3}Nb_{2/3})O_3–xPbTiO_3$ (PZNPT)[2] and $Pb(Mg_{1/3}Nb_{2/3})O_3–xPbTiO_3$ (PMNPT)[3], with morphotropic phase boundaries (MPB). These low symmetry phases are thought to serve as structural bridges between rhombohedral and tetragonal phases and are considered to be responsible for enhanced piezoelectric properties around the MPB compositions because of symmetry-allowed polarization rotation.[4] Three types of ferroelectric monoclinic phases are known, namely, $M_A$, $M_B$ and $M_C$, following the notation of Vanderbilt and Cohen.[5] The former two belong to the space group *Cm* while the latter belongs to *Pm*.[5] For the $M_A/M_B$ phases, the unit cell is double and rotated by 45° about the *c* axis with respect to the pseudocubic cell and the polarization is confined to the $(1\bar{1}0)$ plane. The $M_A$ and $M_B$ unit cells are similar but their magnitudes of the polarization components in the three axes directions corresponding to the pseudocubic unit cell are different: For $M_A$, $P_x = P_y < P_z$, but for $M_B$, $P_x = P_y > P_z$. For the $M_C$ phase, the unit cell is primitive having a unique $b_m$ axis that is oriented along the pseudocubic [010] and the polarization is constrained to lie within the (010) plane. The unit cell parameters for the $M_A$ and $M_C$ phases are shown in Fig. 1(a), which also show their respective polarization vectors denoted as *P*. These two types of monoclinic phases can be distinguished by x-ray reciprocal space mapping (RSM).[6,7] For instance, the 101 reflection in the (H0L) zone splits into two peaks for the $M_A$ phase, whereas it splits into three peaks for the $M_C$ phase, as shown in Fig. 1(b).

Among the lead-free ferroelectrics, $BiFeO_3$ (BFO) has been extensively studied because of its room temperature multiferroicity with potential applications in non-volatile memories, spintronic and piezoelectric devices.[8] At room temperature, bulk BFO exhibits a rhombohedrally distorted perovskite structure, with space group *R3c* and pseudocubic lattice parameters $a_r = 3.965$ Å and $α_r = 89.4°$.[9] The spontaneous polarization of bulk BFO lies along the pseudocubic <111> directions and can reach as high as 100 $μC/cm^2$.[8] For device applications, BFO is preferred in thin film form and is thus subject to a substrate-induced strain. The crystal structure of the epitaxial thin film often deviates from that of the bulk due to strain. For instance, BFO films grown on low misfit substrates reportedly adopt a monoclinic structure of the $M_A$ phase for compressive strain and the $M_B$ phase for tensile strain (space group *Cm* or *Cc*, depending on whether the oxygen octahedra rotations are suppressed by the substrate or not, respectively).[7,10-12] Early first-principles calculations predicted a metastable tetragonal BFO phase (space group *P4mm*) with a giant axial ratio ($a$ = 3.67 Å and $c$ = 4.65 Å) and a large spontaneous polarization (~151 $μC/cm^2$).[13] Subsequent experiments confirmed that such a metastable phase can be stabilized in BFO films by growing on LaAlO$_3$ (LAO) substrates, and suggest that it is not an exact tetragonal *P4mm* but monoclinic with *Cm* or *Cc* symmetry ($M_A$ type).[14-16] Intriguingly, Zeches *et al.*[15] observed MPB-like behavior in mixed-phase films in which the tetragonal-like (T-like) phase coexisted with a rhombohedral-like (R-like) phase. Their first-principles theoretical calculations pointed out that strain-induced phase transition is isosymmetric *Cc-Cc* and the polarization in the T-like phase lies in the ($1\bar{1}0$) plane, but is rotated from [111] direction to nearly [001] direction.[17] However, no contrast was observed in their in-plane PFM images, which is at odds with the *Cc* phase symmetry.[15] To the contrary, several other groups detected in-plane contrast on the T-like domains in films on LAO and LaSrAlO$_4$ substrates.[14,16,18]

To resolve these discrepancies regarding the crystal and domain structure of BFO films grown on LAO substrates, more specifically the nature of the T-like phase, and the so called isosymmetric phase transitions accompanying their formation,[19] in this paper, we embarked on a detailed investigation employing synchrotron x-ray diffractometry and piezoelectric force microscopy (PFM). Our investigations prove that the lattice structure of the T-like phase is the monoclinic $M_C$, instead of the $M_A$ type reported by previous studies.[14-17,20,21] The corresponding ferroelectric domain structure of this $M_C$ phase was also constructed with the aid of PFM images. Additionally, a series of BFO samples with the same film thickness ~70 nm was grown on various substrates to induce misfit strains ranging from tensile to large compressive in order to elucidate the trend in the *c*-lattice parameter and the consequent evolution of the polarization rotation.

Fig. 1 (c) displays a typical $\theta-2\theta$ x-ray diffraction (XRD) pattern of the BFO film on LAO. Intense 00*l* reflections of the T-like phase (*c* = 4.66 ± 0.01 Å) were observed suggesting that this phase can be stabilized by the large compressive misfit strain. Besides, weak peaks corresponding to the 00*l* reflections of the R-like phase (*c* = 3.99 ± 0.01 Å) were also detected, implying the coexistence of T-like and R-like phases. Our XRD result demonstrates that the film is phase pure without any detectable impurity. Fig. 1 (d) shows high resolution synchrotron XRD RSM around the 103 reflection of the T-like phase. It is clear that the 103 reflection splits into three adjacent peaks as a consequence of the existence of four domains: one peak is shifted up and another is shifted down with respect to the central peak. This indicates that the T-like phase must be the monoclinic $M_C$ phase and not the $M_A$ or $M_B$ phases. The monoclinic lattice parameters extracted from the (103) RSM are: $a_m = 3.818$ Å, $b_m = 3.740$ Å, $c_m = 4.662$ Å and $\beta = 88.12°$. This $M_C$ phase is similar to that in PZNPT single crystal near the MPB composition [2] but different from the monoclinic phase ($M_A$-type) reported in BFO films grown on SrTiO$_3$ substrates.[7,10-12]

Piezoelectric force microscopy (PFM) is a tool able to effectively reveal the ferroelectric polarization direction and domain structure of films.[14,22] Fig. 2 (a) shows the topography image of the BFO film on LAO substrate. The film surface is smooth with a root-mean-square (rms) roughness of 10 Å over a 5 μm ×5 μm scan area. The plateau feature arises from the T-like phase while the stripe contrast arises from the multiphase areas as the two phases have different *c*-lattice parameters. Out-of-plane phase (not shown here) image shows uniform contrast, suggesting that all out-of-plane polarizations are pointing in one direction. In contrast to the reports of Bea *et al*.[14] and Zeches *et al*.[15], it was found that the in-plane PFM images of the T-like phase area does exhibit regular in-plane contrast as evident in Fig. 2(b), which is consistent with our previous report on BFO films grown on LaSrAlO$_4$ substrates.[18] Note that three distinct levels of phase contrast are evident in the in-plane PFM image. This observation, together with the uniform contrast shown in the out-of-plane PFM image indicate that the domain structure of the BFO film is characterized by four polarization variants, which is in good agreement with the RSM result. The domain walls here are parallel to the <110> direction, rather than <100> reported for the R-like phase in BFO films on SrTiO$_3$[22] or DyScO$_3$[23] single crystal substrates, which indicates that the T-like phase here should have a different symmetry from that of the R-like phase.

For $M_A$ and $M_C$ phases, the polarization vectors lie in the ($1\bar{1}0$) and (010) planes, respectively.[5] Usually monoclinic symmetry leads to a very complicated domain structure because of many possible domain states.[24] Electric poling or epitaxial strain can reduce the number of domain states to four, allowing for a simplified domain analysis.[2,11,24] The domains with different orientations are separated by domain walls of specified orientations that satisfy the electrical and mechanical boundary conditions. Bokov *et al*.[24] studied the domain structures for the $M_A$ and $M_C$ phases in the poled PMNPT single crystals. Figs. 2 (c) and (d) show the schematic representation of the possible polarization vectors and domain wall traces on the three {100} type planes in these two monoclinic phases. For the monoclinic $M_A$ phase, all permissible uncharged domain walls should intersect the (001) plane in the <100> direction, as shown in Fig. 2(c), which does not agree with our experimental observations in Fig. 2(b). But for the monoclinic $M_C$ phase, the domain walls should intersect the (001) plane in the <110> direction, as shown in Fig. 2(d). The observed domain structure of the T-like phase in our sample shown in Fig. 2(b) fits Bokov's prediction for the $M_C$ phase very well which further supports our suggestion that in our sample the $M_C$ phase is present instead of $M_A$, in consistent with our earlier deduction from the XRD RSM investigation.

A phenomenological analysis has been used to explain the occurrence of $M_C$ phase due to both normal strain and shear strain effects. For (001) oriented BFO single-domain films, a misfit strain-temperature phase diagram has been constructed with different phases: paraelectric $p$ phase with polarization components (0, 0, 0), distorted rhombohedral $r$ phase ($P_x$, $P_x$, $P_z$) i.e. $M_A$ or $M_B$ phase depending on the strain being compressive or tensile, tetragonal $c$ phase (0, 0, $P_z$), and orthorhombic $aa$ phase ($P_x$, $P_x$, 0).[15] We noticed that BFO is deposited at temperatures below its Curie temperature (~ 830 °C) as the ferroelectric rhombohedral phase, which leads to an emergence of shear strain $u_6$ between the film and the substrate. Under this shear strain ($u_6$ = -0.01), our calculation shows the emergence $ac$ phase ($P_x$, 0, $P_z$) and its free energy is close to those of $a$ and $r$ ($M_A$) phases for compressive strain from -0.025 to -0.05, as shown in Fig. 3. For the compressive strain ranging from -0.033 to -0.038 at room temperature shown in inset of Fig. 3, the free energy of $ac$ phase is the lowest compared with those of $c$ and $r$ phases. A similar phenomenon has been reported in PbTiO$_3$ that the shear strain from some non-cubic substrates can suppress the formation of $c$ phase and give rise to $aa$ phase.[25] In the present case, our thermodynamic analysis shows that a strain-induced $ac$ phase in a single domain BFO film may occur at slightly different misfit values with the experimental ones; but qualitatively, this $ac$ phase corresponds to the monoclinic $M_C$ phase with space group $Pm$,[26] which is consistent with the experimentally detected T-like $M_C$ phase in the BFO film on LAO.

In order to understand the evolution of crystal structure and polarization rotation path with changing misfit strain, we carried out a systematic study on the effect of substrate-induced strain on structure of BFO thin films. Epitaxial BFO films were deposited on seven commercially available substrates LAO, NdGaO$_3$, (LaAlO$_3$)$_{0.3}$–(SrAl$_{0.5}$Ta$_{0.5}$O$_3$)$_{0.7}$ (LSAT), SrTiO$_3$, DyScO$_3$, TbScO$_3$, KTaO$_3$ by pulsed laser deposition. All substrates are pseudocubic (001) oriented single crystals. The lattice mismatches between the BFO films and the substrates range from -4.4% for LAO to +0.6% for KTaO$_3$. Fig. 4(a) shows $\theta-2\theta$ XRD scans of the (001) oriented BFO films on different substrates. The 002 peaks attributable to the BFO film are indicated by asterisks. As can be seen, there is a gradual shift in the 002 reflections from KTaO$_3$ to NdGaO$_3$ and finally a relatively larger shift for LAO. The measured $c$-parameter for each film ranges from 3.92 Å to 4.66 Å, as shown in Fig. 4(b). This corresponds to stabilization of $c$-parameter variation of ~20% at a thickness of 70 nm. From previous reports, we infer that for misfit strains from +0.6% to -2.8% (NdGaO$_3$) the R-like phase is obtained which could either be the monoclinic $M_A$ or $M_B$ phase depending on the strain being compressive or tensile, as shown in Fig. 4(b).[10] For the highly strained BFO films such as those on LAO substrate, the T-like phase is predominantly obtained, which is shown to be the monoclinic $M_C$. The sudden shift in the peak position evident in Fig. 4(a) could be attributed to the phase transition $M_A \rightarrow M_C$. Fig. 4(b) also shows out-of-plane lattice strain, $\varepsilon_{zz} = (c - a_r)/a_r$, versus the in-plane misfit strain, $\varepsilon_{xx} = (a_s - a_r)/a_r$ ($a_s$ is the average pseudocubic in-plane lattice parameter of the substrate). Note that the data points for the strained R-like phases ($M_A$ or $M_B$) show a linear relationship between the $c$-parameter (or $\varepsilon_{zz}$) and $\varepsilon_{xx}$, which indicates that the misfit strains can be retained without relaxation and the change of $c$-parameters follows a pure elastic deformation. The single data point for the strained T-like $M_C$ phase obtained in films on LAO is clearly off this linear relationship. This is a further proof that the T-like phase obtained in the BFO film on LAO is different from those strained R-like phase on lower misfit substrates. Poisson ratio $v$ of the strained R-like phases can be calculated using the equation $v = 1/(1-2\varepsilon_{xx}/\varepsilon_{zz})$. $\varepsilon_{xx}/\varepsilon_{zz}$ is estimated by a linear fit (black line) to the data points in Fig. 4(b) which gives the value $v = 0.49 \pm 0.01$. This is so far the largest range (+0.6% to -2.8%) of substrate-induced strain over which Poisson's ratio has ever been determined for BFO films. The obtained value $v \sim 0.5$ is much larger than those previously reported for polycrystalline ($v \sim 0.21$) BFO films[27] and many other perovskite materials ($v = 0.2 \sim 0.4$)[28,29].

Why can the misfit strains be retained without relaxation at a large film thickness (~70nm) and why a large Poisson ratio ~0.5 is obtained for R-like phase? Unlike in most conventional ferroelectrics, the perovskite tolerance factor of BFO is very small (~0.88), which allows for large degrees of the rotation or tilting of oxygen octahedra.[8] Therefore, BFO may use tilting of its oxygen octahedra for strain accommodation at relatively low energy costs.[30] The importance of oxygen octahedra rotation or tilting in BFO has also been emphasized by other reports.[14,31] Furthermore, the crystal structure of strained BFO films is monoclinic, which can provide further degrees of freedom for the polarization rotations. Thus, the energy differences between the different strain states are small, since monoclinic phase is usually accompanied by structural softness.[32] Such structural instability or softness has been hypothesized by a recent first principles study in epitaxial BFO films.[21] Hence, large tetragonal distortion (c/a ratio) can be stabilized even in relatively thicker films with un-relaxed large strains.

The strain-mediated polarization rotation could be inferred from the phase transitions discussed here. Starting from the strain-free rhombohedral (R) phase, the strain-induced transition path is $R \rightarrow M_A$ by compressive strains or $R \rightarrow M_B$ by tensile strains [10]. At large enough compressive strain, the $M_A \rightarrow M_C$ phase transition occurs and brings about a sudden increase in the *c*-lattice parameter, as evident from Fig 4(b), which implies a different polarization rotation path from a simple *Cc-Cc* ($M_A \rightarrow M_A$) phase transition reported in previous studies,[14-17,20,21] as shown in inset of Fig. 4(b). Consequent to this new rotation path, the $M_C$ phase can coexist with the $M_A$ phase at certain misfit strain ranges as shown in Fig. 3. The coexistence of different monoclinic phases $M_C$ and $M_A$ may also allow a variety of other local, distorted phases or domains inside and near the $M_A$ /$M_C$ phase boundaries. This indicates a *soft* lattice for BFO and a tunable behavior by strains where the polarization rotation paths could be mediated in the same way as in those driven by electric field,[2,15] chemical composition,[33] pressure,[34] and temperature[6]. We suggest that first-principles calculations may consider the $M_C$ phase (*Pm* or *Pc* symmetry) and clarify all the possible polarization rotation paths further.

In conclusion, our experimental results suggest that the T-like phase of BFO on LAO is monoclinic $M_C$, which is different from $M_A$ type monoclinic phase reported in films grown on low misfit substrates. A phenomenological analysis used here with both compressive and shear strain effects also suggest the emergence of $M_C$ phase. The *c*-lattice parameter evolution on different substrates further confirmed the deviation of *c*-lattice parameter between the T-like $M_C$ phase and the trend of R-like $M_A$ phase. The presence of this low symmetry phase $M_C$, its linkage to $M_A$ in phase transition regions, and the $M_A$/$M_C$ multiphase coexistence could be one of the major factors that lead to the huge piezoelectric response in BFO films on LAO substrates. The increased variety of domains and phase symmetries on different substrates could add more tunable functionalities associated with enhanced multiferroicity for BFO films mediated by epitaxial strains.

**Experimental**

Epitaxial BFO films were grown on different single crystal substrates by pulsed laser deposition with a KrF excimer laser ($\lambda$=248 nm). The deposition temperature and the oxygen pressure were 700 °C and 100 mTorr, respectively.[18] The thickness of fims, measured by cross-sectional TEM (JEOL 2100F microscope), was around 70 nm. $\theta-2\theta$ x-ray diffraction (XRD) patterns were initially obtained using a four-circle x-ray diffractometer (Panalytical X-pert Pro). High resolution XRD RSM was taken at Shanghai Synchrotron Radiation Facility using beam line 14B1 ($\lambda$ = 1.2378 Å). The RSM is plotted in reciprocal lattice units (r.l.u.) of the LAO substrate (1 r.l.u. = 2π/3.789 Å$^{-1}$). PFM investigations were carried out on an Asylum Research MFP-3D atomic force microscope using Pt/Ir-coated tips.

**Acknowledgement**

We thank Prof. L. Bellaiche and Dr. Wei Ren for valuable discussions. LC acknowledges the support from Nanyang Technological University and Ministry of Education of Singapore under Project No. AcRF RG 21/07 and ARC 16/08. CG thanks the support from the Natural Science Foundation of China (50772106, 50721061), Chinese Academy of Sciences, and the Ministry of Science and Technology of China (2010CB93450004). ZLL thanks the Support from Chinese Universities Scientific Fund. We are grateful to the technical support received at BL14B1 of SSRF.

**Figure Captions**

**Figure 1.** Sketch of the (a) unit cell parameters related to the pseudocubic unit cell and (b) domain configuration in the reciprocal (H0L) plane for the $M_A$ and $M_C$ phases. The red lines represent the directions of the spontaneous polarization. (c) $\theta$-$2\theta$ diffraction pattern of a ~70 nm BFO film on LAO substrate. (d) Synchrotron XRD RSM for the film on LAO near the 103 reflection of the tetragonal-like phase.

**Figure 2.** (a) Topography and (b) in-plane PFM images of the BFO film on LAO substrates. The domain orientation can be extracted from the PFM contrast and orientation of the striped domains. The domain orientation is indicated for four different areas in red small squares. Schematics of the four possible domain states and the elementary motifs formed by permissible domain walls in the monoclinic (c) $M_A$ and (d) $M_C$ phases. Arrows represent the polarization directions.

**Figure 3.** The free energies plotted as a function of misfit strain for different phases at room temperature with a shear strain $u_6$= -0.01. The inset of Fig. 3 is the enlarged section of shaded area. $M_C$ (*ac*) phase is stable when compressive misfit is larger than 3.3%.

**Figure 4.** (a) $\theta$-$2\theta$ scans of BFO films grown on different substrates. The * indicate the BFO peak positions. (b) The out-of-plane lattice parameters (solid red) and lattice strains $\varepsilon_{zz}$ (open blue) plotted as a function of the in-plane misfit strain $\varepsilon_{xx}$. Inset sketches the possible strain-induced rotation path where red arrows represent the polarization directions.

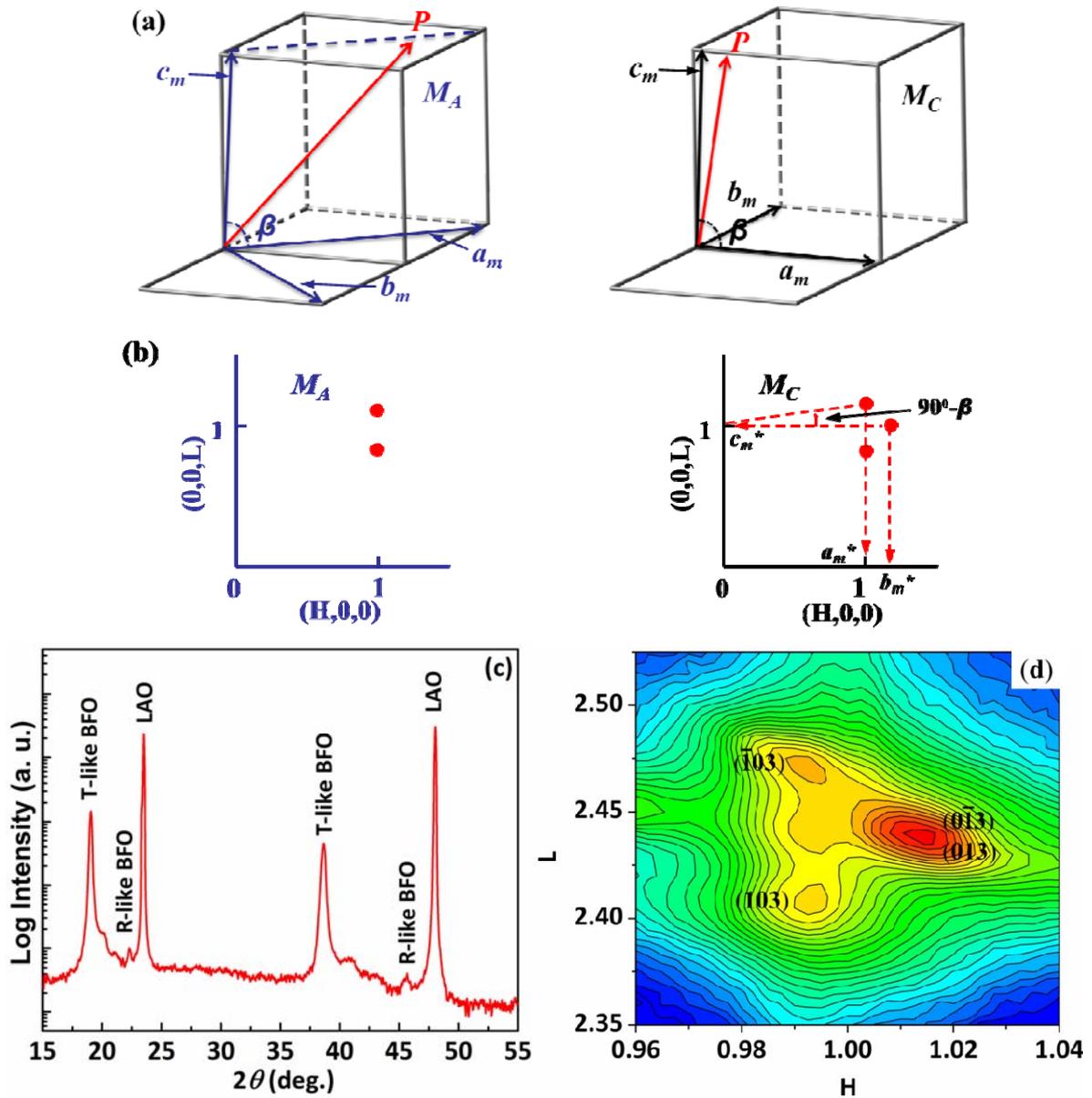

**Figure 1.** Sketch of the (a) unit cell parameters related to the pseudocubic unit cell and (b) domain configuration in the reciprocal (H0L) plane for the $M_A$ and $M_C$ phases. The red lines represent the directions of the spontaneous polarization. (c) $\theta$-$2\theta$ diffraction pattern of a ~70 nm BFO film on LAO substrate. (d) Synchrotron XRD RSM for the film on LAO near the 103 reflection of the tetragonal-like phase.

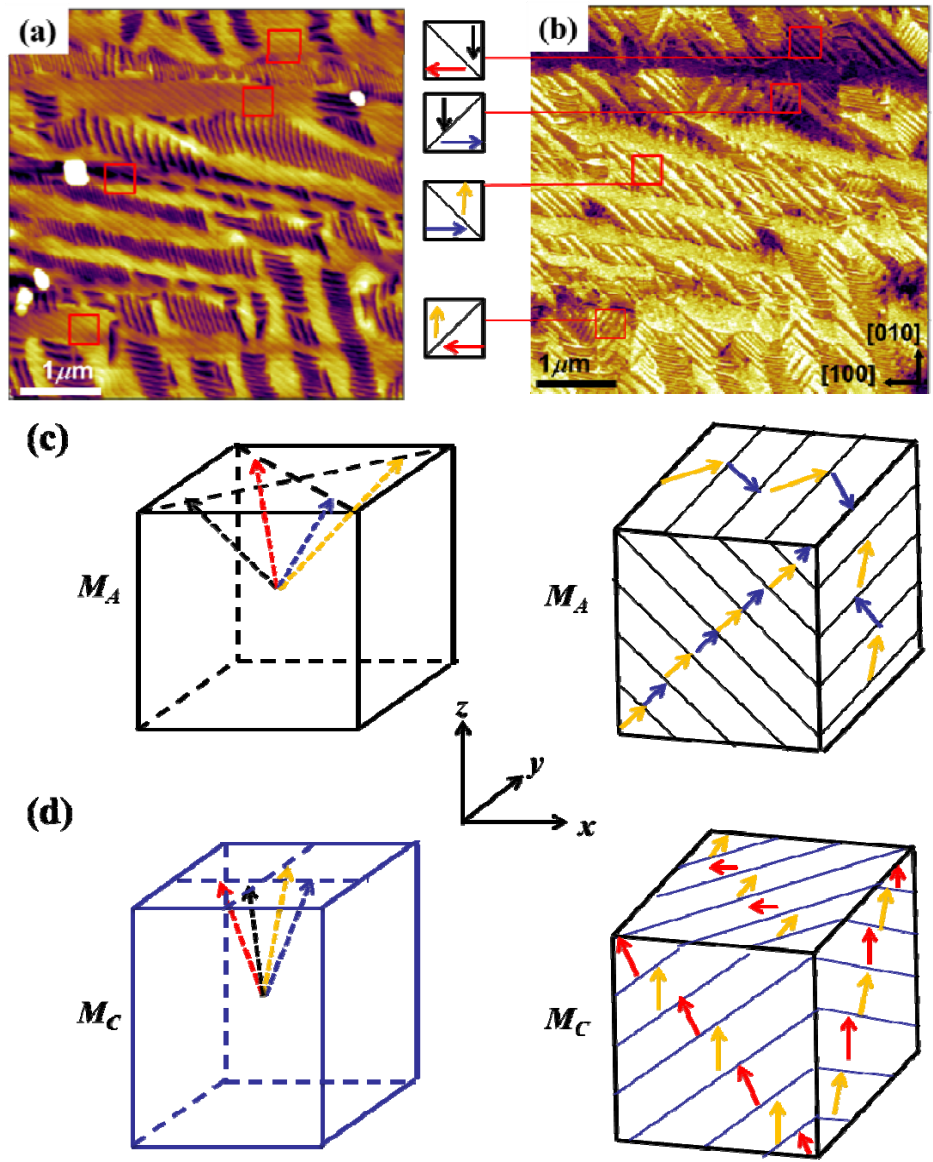

**Figure 2.** (a) Topography and (b) in-plane PFM images of the BFO film on LAO substrates. The domain orientation can be extracted from the PFM contrast and orientation of the striped domains. The domain orientation is indicated for four different areas in red small squares. Schematics of the four possible domain states and the elementary motifs formed by permissible domain walls in the monoclinic (c) $M_A$ and (d) $M_C$ phases. Arrows represent the polarization directions.

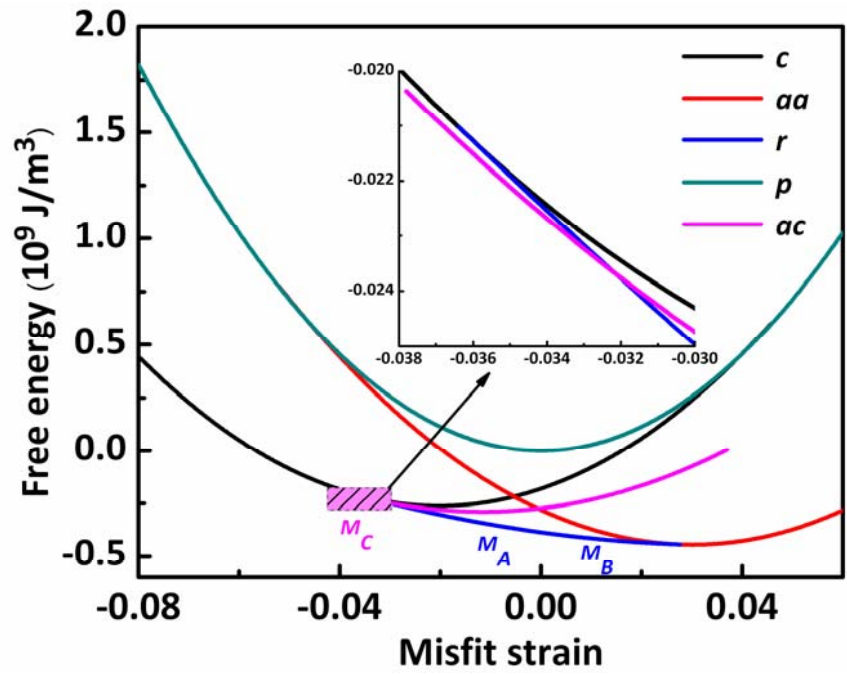

**Figure 3.** The free energies plotted as a function of misfit strain for different phases at room temperature with a shear strain $u_6$ = -0.01. The inset of Fig. 3 is the enlarged section of shaded area. $M_C$ (*ac*) phase is stable when compressive misfit is larger than 3.3%.

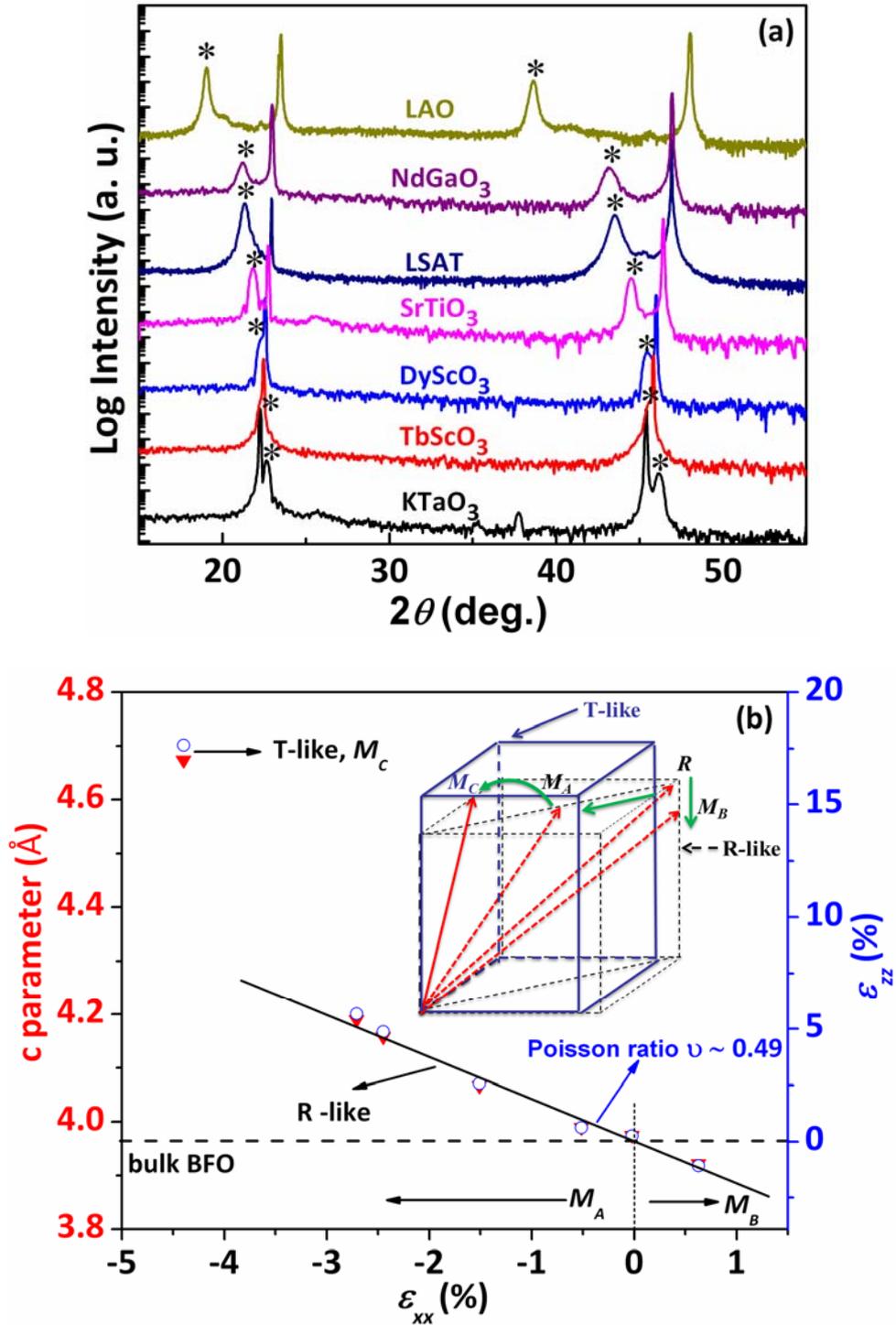

**Figure 4.** (a) *θ-2θ* scans of BFO films grown on different substrates. The * indicate the BFO peak positions. (b) The out-of-plane lattice parameters (solid red) and lattice strains $\varepsilon_{zz}$ (open blue) plotted as a function of the in-plane misfit strain $\varepsilon_{xx}$. Inset sketches the possible strain-induced rotation path where red arrows represent the polarization directions.

**Supporting Information**

**S1. Substrates**

Commercial single crystal substrates bought from CrysTech GmbH. All substrates are (001) oriented single crystals except orthorhombic substrates are (110) oriented. The choice of these substrates is based on their commercial availability and the misfit strains they could produce. The in plane misfit strain $\varepsilon_{xx}$ is defined as:

$$\varepsilon_{xx} = \frac{a_s - a_r}{a_r}$$

where $a_s$ is the average in-plane lattice parameter of the substrate and $a_r$ is the pseudocubic lattice parameter of BFO (3.965Å). By systematically changing the kind of substrate, the strain can be varied. The substrates chosen and the misfit strains they are expected provide to BFO are given in Table S1.

Table S1 Lattice parameters of substrates and in plane misfit strains (negative—compressive strain, positive—tensile strain) of BFO grown on different substrates.

| Material | Structure | Lattice paramter (Å) | | | Misfit strain |
|---|---|---|---|---|---|
| | | a | b | c | |
| LaAlO$_3$ | Rhombohedral | 3.789 | 3.789 | 3.789 | -4.4% |
| NdGaO$_3$ | Orthorhombic | 5.426 | 5.496 | 7.707 | -2.8% |
| LSAT | Cubic | 3.868 | 3.868 | 3.868 | -2.4% |
| SrTiO$_3$ | Cubic | 3.905 | 3.905 | 3.905 | -1.5% |
| DyScO$_3$ | Orthorhombic | 5.440 | 5.713 | 7.890 | -0.5% |
| TbScO$_3$ | Orthorhombic | 5.466 | 5.727 | 7.915 | -0.2% |
| KTaO$_3$ | Cubic | 3.989 | 3.989 | 3.989 | 0.6% |

**S2. Transmission electron microscopy (TEM) imaging of the BFO film on LAO substrate.**

The microstructure of BFO thin films on LAO substrate were examined by using TEM on a JEOL 2100F microscope operated at 200 kV. For preparing cross-sectional TEM specimens, the BFO/LAO sample was cut along the LAO (010) plane, then the slices were thinned by mechanical grinding, polishing, and dimpling, followed by Ar-ion milling at 5 kV. Fig. S1(a) shows the cross-sectional TEM image of the BFO thin film on LAO substrate. The thickness of the film is about 70 nm. Figs. S1(b), (c) and (d) are selected area electron diffraction (SAED) patterns obtained at the positions marked "b", "c" in the BFO film and the LAO substrate, respectively. The SAED patterns from R phase area and LAO substrate are indexed with pseudo-cubic axis. Very flat and sharp interface is seen in the film. The epitaxial nature of the BFO film is evident. The orientation relationships between the film and the substrate can be denoted as (001)BFO//(001)LAO. The SAED pattern in Fig. S1(b) includes both R and T phases, shows the clear splitting of spots indicated by different color points while Fig S1(c) appears to be from a single phase region. Fig. S1(e) shows the HRTEM image obtained from the interface area marked by a rectangle in Fig. S1(a). Fig. S1(e) shows the HRTEM image of BFO and LAO interface from the "e" area marked by a rectangle in Fig. S1(a). Very flat and smooth interface can be seen clearly.

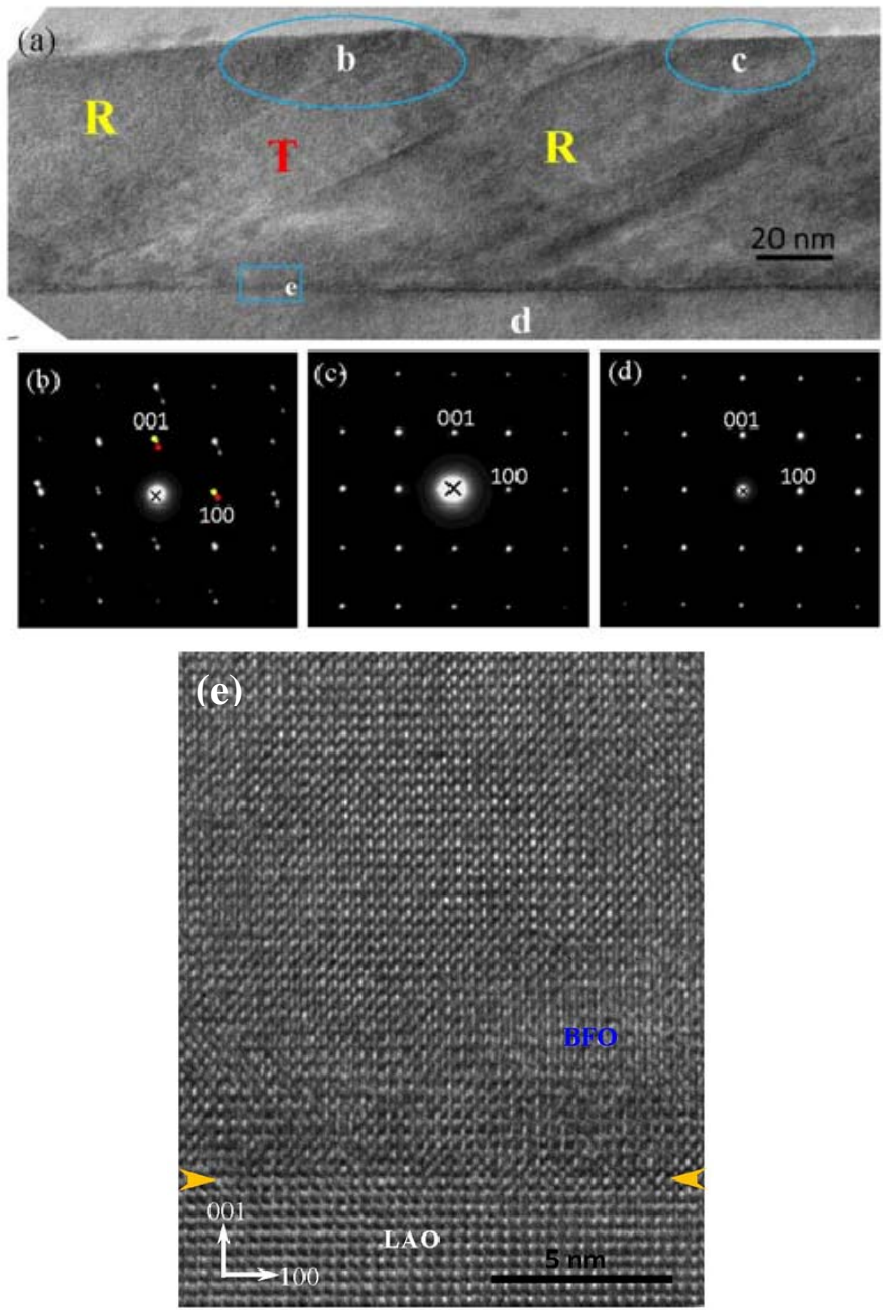

Fig. S1. (a) The cross-sectional TEM image of the BFO thin film on LAO substrate. (b), (c) and (d) are SAED patterns of BFO films from the area marked with "b", "c" and LAO substrate, respectively. The SAED pattern (b) from "b" area includes both R and T phases. (e) HRTEM image of BFO/ LAO interface from the area marked by a rectangle in (a) presented an atomically flat interface.